\documentclass[preprint,aps,superscriptaddress,nofootinbib]{revtex4-1}
\usepackage{graphicx}
\usepackage[toc,page]{appendix}
\usepackage{braket}
\newcommand{\del}{\partial}

\begin{document}

\title{
Spacetime-bridge solutions in vacuum gravity}

\author{Sandipan Sengupta}
\email{sandipan@phy.iitkgp.ernet.in}
\affiliation{Department of Physics and Centre for Theoretical Studies, Indian Institute of Technology Kharagpur, Kharagpur-721302, INDIA}

\begin{abstract}

Spacetimes, which are representations of a bridge-like geometry in gravity theory, are constructed as vacuum solutions to the first order equations of motion.
Each such configuration consists of two copies of an asymptotically flat sheet, connected by a bridge of finite extension where tetrad is noninvertible. 
These solutions can be classified into static and non-static spacetimes.
The associated $SO(3,1)$ invariant fields, namely the metric, affine connection and field-strength tensor, are all continuous
 across the hypersurfaces connecting the invertible and noninvertible phases of tetrad and are finite everywhere. These regular spacetime-bridge solutions do not have any analogue in Einsteinian gravity in vacuum.

\end{abstract}
  
\maketitle

\section{Introduction}
In the study of the global structure of spacetime in general relativity, bridge-like geometries hold a special place due to their rich topology. 
However, Einstein's theory of gravity does not admit such spacetimes as vacuum solutions to the equations of motion. A particularly well-studied example is the Einstein-Rosen bridge \cite{rosen}, which is static and spherically symmetric. This is made up of a pair of asymptotically flat sheets connected at the throat where the area of the spherical slice has a global minimum. 
This configuration satisfies the Einstein's equations everywhere except at the throat where the metric determinant vanishes. This observation had led Einstein and Rosen to propose a modification of the field equations, an approach that was to act as the basis of their remarkable attempt of setting up a geometric model of electrically neutral elementary particles. From the modern perspective though, this configuration is just a part of the maximally extended Schwarzschild geometry in Kruskal coordinates \cite{kruskal}, obtained by deleting the pair of interior (black hole and white hole) regions. This implies that the double-sheeted spacetime of Einstein-Rosen may in fact be envisaged as a non-traversable wormhole \cite{wheeler}. 
Traversable wormholes \cite{ellis,morris} represent another class of bridge geometries.  These typically require the presence of exotic matter (implying violations of the energy conditions \cite{visser}) and hence are not allowed as well within the standard Einsteinian theory.
 
Einsteinian gravity, however, is built upon the assumption that the tetrad (metric) is invertible ($\det e_\mu^I\neq 0$). It represents one among the two possible phases of first order gravity theory, which in general admits invertible as well as noninvertible tetrads ($\det e_\mu^I= 0$)  as spacetime solutions \cite{tseytlin,kaul,kaul1}.
The action associated with the first order formulation is given by the Hilbert-Palatini functional:
 \begin{eqnarray}\label{HP}
S[e,\omega]&=&\frac{1}{8\kappa^2}\int d^4 x~\epsilon^{\mu\nu\alpha\beta}\epsilon_{IJKL}
e_{\mu}^I e_\nu^J R_{\alpha \beta}^{~KL}(\omega)
\end{eqnarray}
The above description involves the tetrad $e_\mu^I(x) $ and the spin-connection $\omega_\mu^{IJ}(x)$ as the independent $SO(3,1)$-valued fields and $R_{\mu \nu}^{~IJ}(\omega)=\del_{[\mu}\omega^{IJ}_{\nu]}+\omega^{IK}_{[\mu}
\omega^{KJ}_{\nu]}$ is the field strength.
The indices $\mu\equiv (t,a)$ label the spacetime coordinates whereas the SO(3,1) indices $I\equiv(0,i)$ correspond to the local Lorentzian frame. The completely antisymmetric tensor densities $\epsilon^{\mu\nu\alpha\beta}$ and $\epsilon_{IJKL}$ take constant values $0,\pm 1$. The above action is to be contrasted with the Einstein-Hilbert (second-order) action, which requires the inverse metric explicitly in its construction and hence can not accomodate any spacetime solution with degenerate tetrads.
The equations of motion in vacuum resulting from the variation of the first order action (\ref{HP}) with respect to the fields $e_\mu^I$ and $\omega_\mu^{IJ}$ are given by, respectively:
\begin{eqnarray}\label{eom1}
e_{[\mu}^{[K} D_{\nu}(\omega)e_{\alpha]}^{L]}&=&0\\
e_{[\nu}^{[J} R_{\alpha\beta]}^{~KL]}(\omega)&=&0\label{eom2}
\end{eqnarray}
As emphasized already, this set of equations admits degenerate as well as nondegenerate spacetime solutions.
In the special case when the tetrad is invertible, these reduce to the Einstein equations of motion, given by: $R_{\mu\nu}(g)=0$. For degenerate tetrads, however, there exists an infinity of solutions, none of which are perceived by the Einsteinian theory \cite{kaul, kaul1}. These spacetimes generically possess torsion, originating due to the noninvertibility of tetrad.

Thus, one is faced with the possibility of exploring a more general dynamical description of pure gravity where the invertible and noninvertible phases of tetrads could coexist. In other words, the equations of motion in first order gravity may exhibit solutions described by a single spacetime, with tetrad fields that are invertible in one region and noninvertible in another \cite{kaul2} (see refs.\cite{bengtsson,madhavan} for a few examples of such spacetime solutions in the context of the complex $SU(2)$ formulation of gravity theory). An interesting question could be, whether such a general framework based on first order gravity could admit vacuum solutions which can be identified as spacetime-bridge geometries (which need not be the same as the Einstein-Rosen bridge or wormholes in particular but may be more generic). Here we demonstrate that it does.

To elaborate further, the vacuum spacetimes presented here consist of an extended bridge-like region, which exhibits a degenerate tetrad and connects two identical sheets of asymptotically flat geometry. There exist two classes of such solutions, namely, static and non-static. For the static ones, the area of the (spatial) two-sphere embedded within the four-geometry has a local maximum at the centre of the bridge. The nonstatic geometries, on the other hand, correspond to  minimum area at their origin. 
 To emphasize, the area of the spherical slice in both these cases is nonvanishing for any arbitrary values of the non-angular coordinates (denoted as $t$ and $u$ here). Hence, these configurations as a whole emerge as a new family of spacetime-bridge geometries in gravity theory, satisfying the first order equations of motion everywhere. From their features outlined above, it is clear that these are not the same as Einstein-Rosen bridge or wormholes in general.
 It is important to note that the $SO(3,1)$ invariant fields that define any of these spacetimes are continuous at the junctions between the invertible and noninvertible phases of the tetrads and are finite everywhere.
 Such regular solutions, being devoid of any matter content (ordinary or exotic), have no analogue in Einsteinian gravity. 

In the next section, we elucidate the method to  construct a family of spacetime-bridge solutions to the  first order equations of motion. First, we present the analysis for regions away from the bridge.  Next, we construct the bridge itself, its geometry being described by noninvertible tetrad fields with one null eigenvalue. Properties such as continuity and finiteness of the basic fields  are also discussed in detail. We conclude with a summary of the essential results and a few relevant remarks.

\section{Generalized wormhole solutions in vacuum}
To begin our analysis, we introduce a set of global spacetime coordinates ($t,v,\theta,\phi$) with $t\in(-\infty,\infty),~v\in(-\infty,\infty),~\theta\in[0,\pi],~\phi\in[0,2\pi]$. In these coordinates, let us divide the whole spacetime into three regions with $-\infty<v<-\epsilon$, $-\epsilon\leq v \leq \epsilon$ and $\epsilon<v<\infty$ for some finite $\epsilon>0$. The metric at the two regions $|v|>\epsilon$ is invertible, whereas at the intermediate region $|v|\leq \epsilon$ it is noninvertible. The asymptotic limits $v\rightarrow \pm\infty$ correspond to the flat spacetime. By glueing the geometries characteristic of the three regions sufficiently smoothly across the (degenerate) phase boundaries $v=\pm \epsilon$, one can obtain the  full spacetime solution which satisfies the first order equations of motion (\ref{eom1}) and (\ref{eom2}) everywhere (i.e. for $-\infty<v<\infty$).

\subsection{Regions away from the bridge: Invertible tetrad}
We assume the geometry of each of the identical sheets $\epsilon<|v|<\infty$ to be described by a static metric, given by:
\begin{eqnarray}\label{sc}
ds^2~=~g_{\rho\sigma}dx^{\rho}dx^{\sigma}~=~&-&\left[\frac{f^2(v)}{f^2(v)+2M}\right]dt^2 + 4\left[f^2(v)+2M\right]f^{'2}(v) dv^2\nonumber\\ &+&~ \left[f^2(v)+2M\right]^2\left[d\theta^2+\mathrm{sin}^2 \theta d\phi^2\right]
\end{eqnarray}
The monotonic function $f(v)$ above has the following behaviour at the junctions $v=\pm \epsilon$ and the asymptotic boundaries:
\begin{eqnarray}\label{f(u)}
f(\pm\epsilon)=0=f'(\pm\epsilon)~,~f(v)\rightarrow \infty \mathrm{~as~}v\rightarrow \infty~.
\end{eqnarray}
The last condition in eq.(\ref{f(u)}) implies that the metric (\ref{sc}) is asymptotically flat.
The constant $M$ defines the area $A~(=16\pi M^2)$ of either of the two-spheres at $t=const.,~v=\pm \epsilon$. Upto the boundary conditions (\ref{f(u)}), $f(v)$ can be any arbitrary function, as long as it does not lead to any divergence in the metric, connection or field strength tensor.
The corresponding tetrad fields read:
\begin{eqnarray}\label{tetrad}
e^0&=&\frac{f(v)}{[f^2(v)+2M]^{\frac{1}{2}}} dt,~ e^1=2[f^2(v)+2M]^{\frac{1}{2}}f'(v)dv,\nonumber\\
e^2&=&[f^2(v)+2M]d\theta,~e^3=[f^2(v)+2M]\sin\theta d\phi
\end{eqnarray} 
Evidently, the tetrad is invertible everywhere at $|v|>\epsilon$ but not at the phase boundaries $v=\pm \epsilon$. Note that in the regions $|v|>\epsilon$, the metric above can be brought to the 
Einstein-Rosen form \cite{rosen} using the coordinate transformation $f(v)=u$. However, the equivalence between the two geometries breaks down at $v=\pm \epsilon$ where the Jacobian of the transformation is singular.

The nonvanishing components of the torsionless spin-connection fields $\omega_\mu^{~IJ}(e)$ are:
\begin{eqnarray}\label{omega}
\omega_t^{01}=\frac{M}{[f^2(v)+2M]^2},~\omega_\theta^{12}=-\frac{f(v)}{[f^2(v)+2M]^{\frac{1}{2}}},~\omega_\phi^{23}=-\cos\theta,~
\omega_\phi^{31}=\frac{f(v)}{[f^2(v)+2M]^{\frac{1}{2}}} \sin\theta~~~
\end{eqnarray}
Evaluation of the field strength $R^{~IJ}_{\mu\nu}(\omega)$ using these leads to the following nontrivial components:
\begin{eqnarray}\label{R}
R^{01}(\omega)&=&-\frac{4M f(v)f'(v)}{[f^2(v)+2M]^{3}}dt\wedge dv,~~R^{02}(\omega)=-\frac{M f(v)}{[f^2(v)+2M]^{\frac{5}{2}}}dt\wedge d\theta,\\R^{03}(\omega)&=&-\frac{M f(v)}{[f^2(v)+2M]^{\frac{5}{2}}}\sin\theta dt\wedge d\phi,~~
R^{12}(\omega)=-\frac{2M f'(v)}{[f^2(v)+2M]^{\frac{3}{2}}} dv\wedge d\theta,\\R^{23}(\omega)&=&
\frac{2M }{[f^2(v)+2M]}\sin\theta d\theta\wedge d\phi,~~R^{31}(\omega)=-\frac{2M f(v)}{[f^2(v)+2M]^{\frac{3}{2}}}\sin\theta d\phi\wedge dv
\end{eqnarray}

From these, we can construct another set of variables, namely the affine connection $\Gamma_{\alpha\beta\rho}$ and the field-strength $R_{\alpha\beta\rho\sigma}$, which are  invariant under the internal $SO(3,1)$ rotations. The first is defined through the covariant constancy of the metric, given by the condition ${\cal D}_{\alpha} g_{\rho\sigma}\equiv \del_{\alpha}g_{\rho\sigma}-\Gamma_{\alpha\rho\sigma}-\Gamma_{\alpha\sigma\rho}=0$, implying:
\begin{eqnarray}\label{gamma1}
\Gamma_{\alpha\beta\rho}=\frac{1}{2}\left[\del_{\alpha} g_{\beta\rho}+  \del_{\beta} g_{\alpha\rho}-  \del_{\rho}g_{\alpha\beta}\right]-K_{\alpha\beta\rho}
\end{eqnarray}
where $K_{\mu\alpha\nu}=-K_{\mu\nu\alpha}=\frac{1}{2}[\Gamma_{[\mu\nu]\alpha}+\Gamma_{[\alpha\mu]\nu}+\Gamma_{[\alpha\nu]\mu}]$ is the contortion, defined in terms of the antisymmetric part $\Gamma_{[\mu\nu]\alpha}=\Gamma_{\mu\nu\alpha}-\Gamma_{\nu\mu\alpha}$ of the affine connection. Since the regions $|v|>\epsilon$ are torsion-free ($K_{\mu\nu\alpha}=0$), the nonvanishing connection components are found to be:
\begin{eqnarray}\label{gamma2}
&&\Gamma_{ttv}=\frac{2Mf(v)f'(v)}{\left[f^2(v)+2M\right]^2},~\Gamma_{tvt}=-\frac{2Mf(v)f'(v)}{\left[f^2(v)+2M\right]^2}=\Gamma_{vtt},~\Gamma_{vvv}=2\del_v \left(\left[f^2(v)+2M\right]f^{'2}(v)\right),\nonumber\\
&&\Gamma_{\theta\theta v}=-2f(v)f'(v)\left[f^2(v)+2M\right],~\Gamma_{\phi\phi v}=-2f(v)f'(v)\left[f^2(v)+2M\right] \sin^2 \theta,\nonumber\\
&&\Gamma_{v\theta\theta}=2f(v)f'(v)\left[f^2(v)+2M\right]=\Gamma_{\theta v \theta},~\Gamma_{v\phi\phi}=2f(v)f'(v)\left[f^2(v)+2M\right]\sin^2 \theta=\Gamma_{\phi v\phi},\nonumber\\
&&\Gamma_{\phi\phi\theta}=-\left[f^2(v)+2M\right]^2 \sin\theta\cos\theta,~\Gamma_{\theta\phi\phi}=\left[f^2(v)+2M\right]^2 \sin\theta \cos\theta=\Gamma_{\phi\theta\phi}~
\end{eqnarray}
The spacetime field-strength tensor $R_{\mu\nu\rho\sigma}$, defined as:
\begin{eqnarray}
R_{\mu\nu\rho\sigma}=R_{\mu \nu}^{~IJ}(\omega) e_{\rho I} e_{\sigma J},
\end{eqnarray}
exhibits the following nonvanishing components:
\begin{eqnarray}\label{riemann1}
R_{tvtv}=-\frac{8Mf^2(v)f^{'2}(v)}{\left[f^2(v)+2M\right]^3},~R_{t\theta t\theta}=\frac{Mf^2(v)}{\left[f^2(v)+2M\right]^2},~R_{t\phi t\phi}=\frac{Mf^2(v)}{\left[f^2(v)+2M\right]^2} \sin^2 \theta,\nonumber\\
R_{v\theta v\theta}=-4Mf^{'2}(v),~R_{\theta\phi \theta\phi}=2M\left[f^2(v)+2M\right]\sin^2 \theta,~R_{\phi u \phi u}=-4Mf^{'2}(v) \sin^2 \theta ~
\end{eqnarray}
The configuration ($e_\mu^I,\omega_\mu^{IJ}$) at $|v|>\epsilon$, defined completely by eqs.(\ref{tetrad}) and (\ref{omega}), satisfies the Einstein equations in vacuum, which describes the invertible phase of first order gravity theory.  
Let us also note that at the hypersurfaces $v=\pm\epsilon$, all the components of the affine connection and field-strength, except the ones below, vanish:
\begin{eqnarray}\label{R2}
&& \Gamma_{\theta\phi\phi}= \Gamma_{\phi\theta\phi}\doteq 4M^2 \sin\theta\cos\theta,~\Gamma_{\phi\phi\theta}\doteq -4M^2 \sin\theta\cos\theta;\nonumber\\
&& R_{\theta\phi\theta\phi}\doteq 4M^2 \sin^2 \theta~
\end{eqnarray}
where $\doteq$ denotes equality only at $v=\pm \epsilon$. 

Although it is not essential to choose an explicit form of $f(v)$ in order to set up the subsequent analysis, we shall do that hereon for definiteness: 
\begin{eqnarray}\label{f1}
f(v)=(v^2-\epsilon^2)^{n},
\end{eqnarray}
where $n>1$ is an integer, a criterion that is necessary for the finiteness of the fields and their derivatives upon continuation to $v=\pm \epsilon$.
It is straightforward to check that this choice is consistent with the boundary conditions (\ref{f(u)}) and that all the associated fields which depend on $f(v)$ or its derivatives are finite for any finite range of $v$ throughout the two regions at $|v|>\epsilon$.

\subsection{Region within the bridge: Noninvertible tetrad}
For the intermediate region $-\epsilon\leq v\leq\epsilon$, which defines an extended bridge between the asymptotically flat sheets, we shall construct a degenerate spacetime solution of the first order equations of motion (\ref{eom1}) and (\ref{eom2}). In particular, we assume that the geometry here is described by a metric with one zero eigenvalue ($\hat{g}_{tt}=0$):
\begin{eqnarray}\label{G}
\hat{ds}^2_{(4)}~\equiv~ \hat{g}_{\rho\sigma}dx^{\rho}dx^{\sigma}~=~0+\sigma F^2(v) dv^2 + H^2(v)\left[d\theta^2+\mathrm{sin}^2 \theta d\phi^2\right]
\end{eqnarray}
Here $F(v)$ and $H(v)$ are two arbitrary functions to be solved using the equations of motion and $\sigma=\pm 1$. The internal  metric is Lorentzianin within the bridge as well, being defined as$\eta_{IJ}\equiv diag[-\sigma,\sigma,1,1,]$. The tetrad fields read:
\begin{eqnarray}\label{e}
\hat{e}^I_\mu~=~\left(\begin{array}{cccc}
0 & 0 & 0 & 0\\
0 & F(v) & 0 & 0\\
0 & 0 & H(v) & 0\\
 0 & 0 & 0 & H(v)\sin\theta  \end{array}\right)  ~=~\left(\begin{array}{cc}
0 & 0 \\
0 & \hat{e}_a^i \end{array}\right) 
\end{eqnarray}
where $\hat{e}_a^i$ are the triads associated with the nondegenerate 3-geometry in metric (\ref{G}). In what follows next, we shall find a set of spin-connection fields such that the pair $(\hat{e}_\mu^I,\hat{\omega}_\mu^{IJ})$ represents a regular solution of the first-order equations of motion (\ref{eom1}) and (\ref{eom2}) everywhere at $-\epsilon\leq v\leq \epsilon$ and also satisfies the continuity requirements at the junctions $v=\pm \epsilon$. 

For a degenerate metric with $\hat{e}_t^I=0=\hat{e}_a^0$ as above, the most general solution of the connection equations of motion (\ref{eom1}) is given by \cite{kaul}:
\begin{eqnarray}\label{omega*}
\hat{\omega}_t^{~0i}=0,~\hat{\omega}_t^{~ij}=0,~\hat{\omega}_a^{~0i}=\epsilon^{ijk}\hat{e}_a^l M_{kl},~\hat{\omega}_a^{~ij}=\bar{\omega}_a^{~ij}(\hat{e})+\epsilon^{ijk}\hat{e}_a^l N_{kl}~,
\end{eqnarray}
where $M_{kl}=M_{lk}$ and $N_{kl}=N_{lk}$ are two arbitrary spacetime dependent symmetric $3\times 3$ matrices and $\bar{\omega}_a^{~ij}(\hat{e})=\frac{1}{2}\left[\hat{e}^b_i\del^{}_{[a}\hat{e}_{b]}^j
-\hat{e}^b_j\del^{}_{[a}\hat{e}_{b]}^i -  \hat{e}_a^l \hat{e}^b_i \hat{e}^c_j
\del^{}_{[b}\hat{e}_{c]}^l\right]$ are the torsionless spin-connection fields  determined by the triads. The equation above  implies that when tetrad is not invertible, the connection equations (\ref{eom1}) do not solve all the components of the $SO(3,1)$ gauge field $\hat{\omega}_\mu^{~IJ}$ completely in terms of the tetrad.  The arbitrary fields $(M_{kl},N_{kl})$ above represent precisely the twelve connection components that are left undetermined. These fields lead to nonvanishing torsion in the theory, whose origin is purely geometric since there is no matter coupling. This is in contrast to the case of invertible tetrads where the connection equations can be solved for all the connection components in terms of the tetrad as $\omega_\mu^{~IJ}=\omega_\mu^{~IJ}(e)$ and torsion vanishes as a consequence.

Just for simplicity, we shall now adopt an ansatz which allows us to work with only one contortion field instead of twelve: 
\begin{eqnarray}\label{N}
  M^{kl}(t,v,\theta,\phi)=0,~~N^{kl}(t,v,\theta,\phi)~=~\left(\begin{array}{ccc}
0 & 0 & 0\\
0 & 0 & \mu(v)\\
0 & \mu(v) & 0 \end{array}\right) 
\end{eqnarray}
A more general analysis with more number of fields is straightforward and does not add much to the essential details that concern us here.
The ansatz (\ref{N}) implies that the contortion one-forms can be written as:
\begin{eqnarray}
K^{12}=\mu(v)H(v) d\theta,~K^{23}=0,~K^{31}=\mu(v)H(v)\sin\theta d\phi~,
\end{eqnarray}
leading to the following expression for the (full) spin-connection fields:
\begin{eqnarray}\label{omegafull}
&&\hat{\omega}^{0i}=0,~
\hat{\omega}^{12}=\left[\mu(v) H(v)-\sigma\frac{H'(v)}{F(v)}\right]d\theta,~
\hat{\omega}^{23}=-\cos\theta d\phi,\nonumber\\
&&\hat{\omega}^{31}=\left[\mu(v) H(v)+\sigma\frac{H'(v)}{F(v)}\right]\sin\theta d\phi
\end{eqnarray}
The curvature two-forms, which are completely determined by the above, read:
\begin{eqnarray}\label{R-simple}
\hat{R}^{0i} (\hat{\omega})&=& 0~,~~\hat{R}^{12}(\hat{\omega})=\left(\mu(v) H(v)-\sigma\frac{H'(v)}{F(v)}\right)' dv\wedge d\theta ~,\nonumber\\
\hat{R}^{23}(\hat{\omega})&=&\left[1+\sigma\left(\mu(v) H(v)-\frac{H'(v)}{F(v)}\right)\left(\mu(v) H(v)+\frac{H'(v)}{F(v)}\right)\right]\sin\theta ~d\theta\wedge d\phi ~,\nonumber\\
\hat{R}^{31}(\hat{\omega})&=&-\left(\mu(v) H(v)+\sigma\frac{H'(v)}{F(v)}\right)'\sin\theta ~d\phi\wedge dv~+~2\mu(v) H(v) \cos\theta~ d\theta\wedge d\phi~.
\end{eqnarray}
The $SO(3,1)$ invariant counterparts of the above fields are now given by the affine connection $\hat{\Gamma}_{\alpha\beta\rho}$ and field-strength tensor $\hat{R}_{\alpha\beta\rho\sigma}$ (with torsion), defined as:
\begin{eqnarray}\label{gamma1}
\hat{\Gamma}_{\alpha\beta\rho}&=&\frac{1}{2}\left[\del_{\alpha} \hat{g}_{\beta\rho}+  \del_{\beta} \hat{g}_{\alpha\rho}-  \del_{\rho}\hat{g}_{\alpha\beta}\right]-K_{\alpha\beta\rho}\nonumber~,\\
\hat{R}_{\mu\nu\rho\sigma}&=&\hat{R}_{\mu \nu}^{~IJ}(\omega) e_{\rho I} e_{\sigma J}~.
\end{eqnarray}
 The nontrivial components of these fields are listed below:
\begin{eqnarray}\label{gamma4}
&&\hat{\Gamma}_{vvv}=\frac{\sigma}{2}\del_v F^2 (v),~\hat{\Gamma}_{\theta\theta v}=-\frac{1}{2}\del_v H^2 (v)+\sigma \mu(v)F(v)H^2(v),\nonumber\\
&&\hat{\Gamma}_{\phi\phi v}=-\left[\frac{1}{2}\del_v H^2(v)+\sigma \mu(v)F(v)H^2(v) \right]\sin^2 \theta,~\hat{\Gamma}_{v\theta\theta}=\frac{1}{2}\del_v H^2(v)=\hat{\Gamma}_{\theta v \theta},\nonumber\\
&&\hat{\Gamma}_{\phi\phi\theta}=-H^2 (v) \sin\theta\cos\theta,~\hat{\Gamma}_{\theta\phi\phi}=H^2(v) \sin\theta \cos\theta=\hat{\Gamma}_{\phi\theta\phi}~;\nonumber\\
&& \hat{R}_{v\theta v\theta}=\sigma\left(\mu(v) H(v)-\sigma\frac{H'(v)}{F(v)}\right)F(v)H(v),\nonumber\\
&& \hat{R}_{\theta\phi \theta\phi}=\left[1+\sigma\left(\mu(v) H(v)-\frac{H'(v)}{F(v)}\right)\left(\mu(v) H(v)+\frac{H'(v)}{F(v)}\right)\right]H^2(v) \sin^2 \theta,\nonumber\\
&& \hat{R}_{\phi u \phi u}=- \sigma\del_v \left(\mu(v) H(v)+\sigma\frac{H'(v)}{F(v)}\right)F(v)H(v)\sin^2 \theta ~.
\end{eqnarray}  

The equations of motion (\ref{eom2}), which remain to be solved, implies \cite{kaul}:
\begin{eqnarray*} 
\epsilon^{abc}\epsilon_{ijk}\hat{e}_a^i \hat{R}_{bc}^{~jk}=0
\end{eqnarray*} 
 Using the expressions given above,  this reduces to the following constraint in terms of the unknown fields $H(v),F(v)$ and $\mu(v)$:
\begin{eqnarray}\label{master1}
\left(\sigma+\mu^2(v) H^2(v)
-\frac{H'^2(v)}{F^2(v)}\right)F(v)-2 H(v)\left(\frac{H'(v)}{F(v)}\right)'=0
~
\end{eqnarray}
Since there are no more equations of motion to be solved, we must choose two further conditions in order to obtain explicit solutions for these three fields. Further, these constraints must be consistent with the boundary conditions at the phase boundaries $ v=\pm \epsilon$. A possible choice is:
\begin{eqnarray}\label{MC}
\mu(v)H(v)=\sigma\frac{H'(v)}{F(v)},~
\frac{H'(v)}{F(v)}=\lambda (v^2-\epsilon^2)^m
\end{eqnarray} 
where $m>1$ is an integer and $\lambda$ is a constant.
The three equations in (\ref{master1}) and (\ref{MC}) can now be solved for $H(v),F(v)$ and $\mu(v)$, leading to:
\begin{eqnarray}\label{nonsingular2}
F(v)&=&8\lambda Mm v (v^2-\epsilon^2)^{m-1} e^{\sigma\lambda^2(v^2-\epsilon^2)^{2m}},  \nonumber\\ 
H(v)&=&H_0 e^{\sigma\lambda^2 (v^2-\epsilon^2)^{2m}},
\nonumber\\ 
\mu(v)&=& \sigma\frac{\lambda}{H_0} (v^2-\epsilon^2)^m e^{-\sigma\lambda^2(v^2-\epsilon^2)^{2m}}~.
\end{eqnarray}
Continuity of the metric at $v=\pm \epsilon$ fixes the constants $H_0,\lambda$ and the exponent $m$ as:
\begin{eqnarray}\label{cont}
 \lambda^2=\frac{\sigma}{2M},~H_0=2M,~m=n,
\end{eqnarray}
where $n$ is the same integer that defines the function $f(v)$ in (\ref{f1}) at the regions $|v|>\epsilon$. 

Let us now summarize some of the important features of the solutions just obtained. 
The first in the set of equations (\ref{cont}) implies that there exist two classes (static and non-static) of solutions, corresponding to $\sigma=+1,~M>0$ and $\sigma=-1,~M<0$, respectively. In these two cases, $v$ within the bridge behaves like a spacelike and timelike coordinate, respectively:
\newpage
\begin{eqnarray}\label{nonsingular3}
\sigma &=&+1:\nonumber\\
&& F(v)= (32M)^{\frac{1}{2}} n v (v^2-\epsilon^2)^{n-1} e^{\left[\frac{1}{2M}(v^2-\epsilon^2)^{2n}\right]},  \nonumber\\ 
&& H(v)=2M e^{\left[\frac{1}{2M} (v^2-\epsilon^2)^{2n}\right]},\nonumber\\ 
&& \mu(v)= (2M)^{-\frac{3}{2}} (v^2-\epsilon^2)^n e^{\left[-\frac{1}{2M}(v^2-\epsilon^2)^{2n}\right]}~\nonumber\\
\sigma &=&-1:\nonumber\\
&& F(v)= (32M)^{\frac{1}{2}} n v (v^2-\epsilon^2)^{n-1} e^{\left[-\frac{1}{2M}(v^2-\epsilon^2)^{2n}\right]},  \nonumber\\ 
&& H(v)=2M e^{\left[-\frac{1}{2M} (v^2-\epsilon^2)^{2n}\right]},\nonumber\\ 
&& \mu(v)= -(2M)^{-\frac{3}{2}} (v^2-\epsilon^2)^n e^{\left[\frac{1}{2M}(v^2-\epsilon^2)^{2n}\right]}~,
\end{eqnarray}
For the static solutions, the nondegenerate three-space exhibits a bridge-topology. For the non-static class, however, the bridge resides within the Lorentzian three-geometry. 
Snapshots of these two classes of solutions (at a fixed t) are provided in FIG.1, where each circle represents a two-sphere covered by the angles $(\theta,\phi)$. 
\begin{figure}\begin{center}
\includegraphics[height=13.5cm]{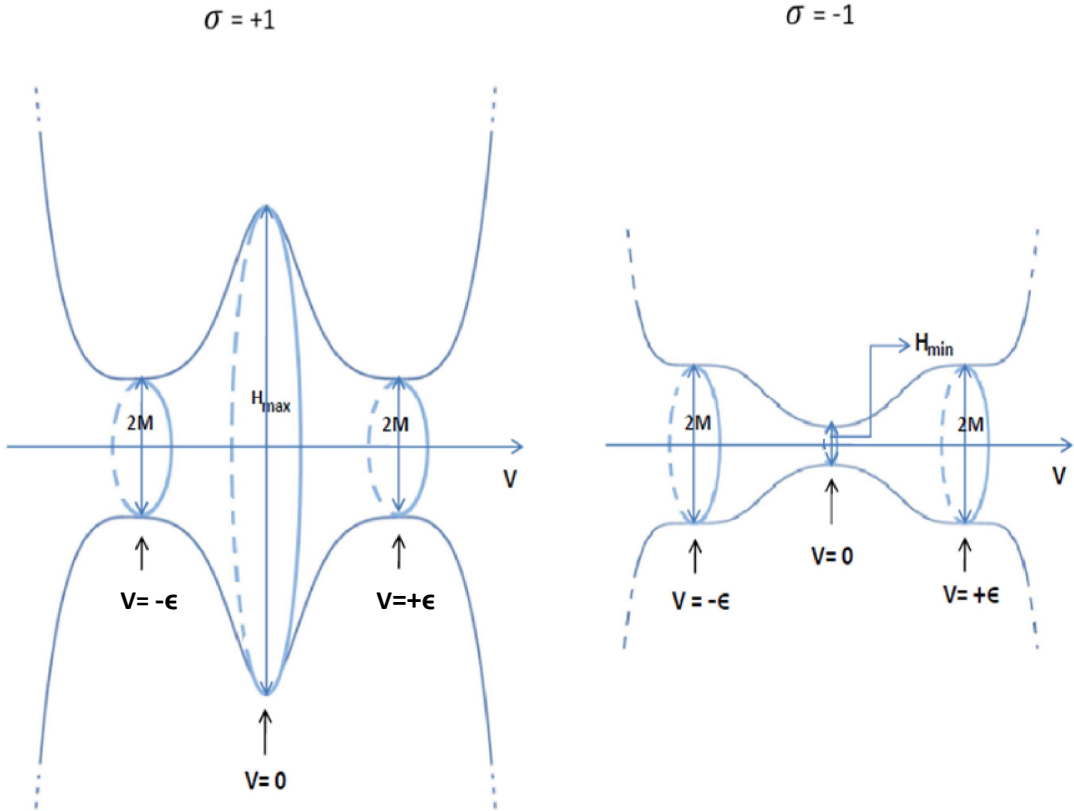}
\caption{Representation of spacetime solutions at a fixed $t$ as bridge-like three-geometries}
\end{center}
\end{figure}Each solution (for a fixed integer $n=m$) is associated with two free parameters $\epsilon$ and $M$. While $\epsilon$ defines the size of the degenerate bridge along the $v$ direction, $M$ characterizes the area of the two-spheres at its boundaries $v=\pm \epsilon$. 
At the origin $v=0$, the radius $H(v)$ of the two-sphere corresponds to a (local)  maximum for $\sigma=1$ and a (global) minimum for $\sigma=-1$, with:
\begin{eqnarray}\label{alpha}
H_{max}&=&2M e^{\left[\frac{\epsilon^{4n}}{2M}\right]}>2M \mathrm{ ~for ~\sigma=+1},\nonumber\\
H_{min}&=&2M e^{\left[-\frac{\epsilon^{4n}}{2M}\right]}<2M \mathrm{~ for ~\sigma=-1}~.
\end{eqnarray}
It is worth emphasizing that for any of the non-static solutions with $\sigma=-1$, the minimum radius $H_{min}$ is less than $2M$.
The profile of the contortion field $\mu(v)$, which is completely localized within the  bridge, is given by FIG.2.
\begin{figure}\begin{center}
\includegraphics[height=11cm]{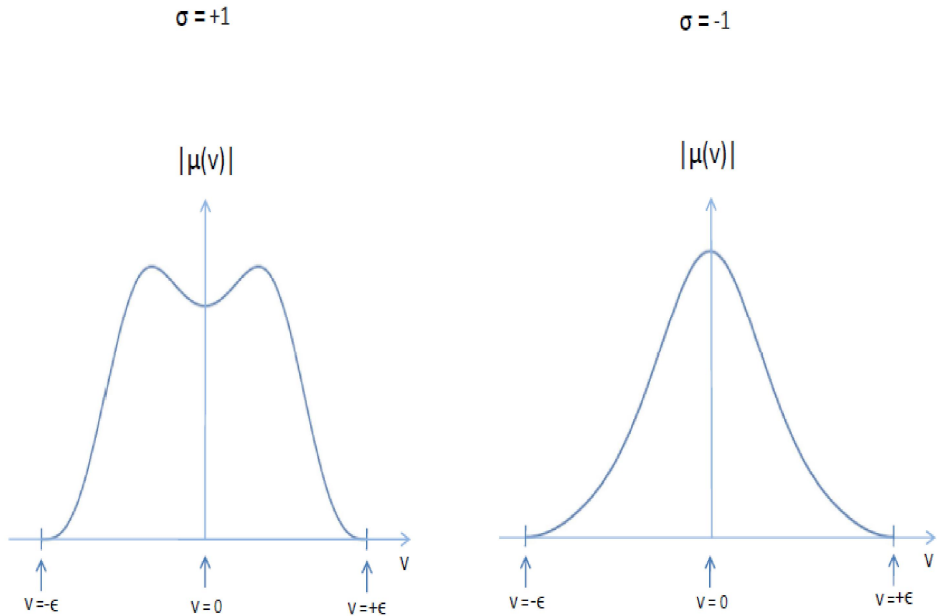}
\caption{Profile of contortion field $\mu(v)$}
\end{center}
\end{figure}

The solutions obtained above, when inserted into the expressions for the affine connection and field-strength in eq.(\ref{gamma4}), imply that the continuity requirements at the phase boundaries $v=\pm \epsilon$ are all satisfied:
\begin{eqnarray*}
 g_{\mu\nu}\doteq \hat{g}_{\mu\nu},~ \Gamma_{\alpha\beta\rho}\doteq\hat{\Gamma}_{\alpha\beta\rho},~R_{\alpha\beta\rho\sigma}\doteq~\hat{R}_{\alpha\beta\rho\sigma}.
\end{eqnarray*} 
Note that $g_{\mu\nu}(u)$ is a $C^{2n-1}$ function at these junctions.

This completes the construction of the spacetime-bridge solutions of the first order equations of motion in vacuum, given by eqs. (\ref{eom1}) and (\ref{eom2}). For the static as well as non-static geometries, the ($t=const.$) spherical slices have nonvanishing radii for any arbitrary value of the coordinates $v$, a feature that epitomizes the bridge topology of these solutions. 
The bridge, defined by degenerate tetrad and torsionful connection fields, interpolates between two asymptotically flat spacetime sheets. The full spacetime  is continuous and the associated fields, in particular the metric, connection and field-strength, are finite everywhere. 

Although we have used a particular form of the contortion matrix $N_{kl}$ in eq.(\ref{N}) (with a pair of non-diagonal entries being non-zero) to obtain the solutions displayed here, that is not really necessary. For instance, the same set of solutions may be obtained using a matrix whose only nontrivial elements are  diagonal instead:
\begin{eqnarray}
 N^{kl}(t,v,\theta,\phi)~=~\left(\begin{array}{ccc}
\alpha(v) & 0 & 0\\
0 & \beta(v) & 0\\
0 & 0 & 0 \end{array}\right) 
\end{eqnarray}
where $\alpha(v)=-\sigma \beta(v)$. In this case, the contortion field $\alpha(v)$ would play the role of $\mu(v)$ as appearing in the solutions (\ref{nonsingular3}) for $\sigma=\pm 1$.

Note that within this framework, the two-sheeted spacetime constructed by Einstein-Rosen \cite{rosen} emerges as a special case. This corresponds to the limit $\epsilon\rightarrow 0$ (with M kept fixed) of the static solutions with $\sigma=+ 1,~M>0$. For this geometry, torsion vanishes everywhere, including at the hypersurface $v=0$ where the pair of degenerate phase boundaries coincide. 

\section{Conclusions}
First order formulation of gravity theory is known to admit two possible phases associated with invertible and noninvertible tetrads. While the invertible phase is equivalent to Einstein's theory gravity, the other is not. Here we have demonstrated that when the two phases coexist, the theory admits a new class of vacuum solutions which are representations of a spacetime-bridge geometry. The bridge, which connects a pair of asymptotically flat sheets, is defined by noninvertible tetrad and has a finite extension. Away from the bridge, the tetrad fields are invertible.
As one approaches the bridge from any of the outer regions, the determinant of the four-metric goes to zero continuously at the phase boundaries. 
The fact that first order gravity admits such nontrivial (bridge) topologies as regular vacuum solutions is in stark contrast to the case of Einsteinian theory.

Each of the countable infinity of solutions 
 exhibits two free parameters $M$ and $\epsilon$, which essentially define the area and the location of the junctions between the invertible and noninvertible phases of tetrad. 
Solutions exist for both the signs of M, leading to their classification into static ($M>0$) and non-static ($M<0$) spacetimes. Within the framework set up here, the Einstein-Rosen configuration emerges as a special limit of the static configurations with vanishing torsion. 
 It is obvious that being devoid of matter (ordinary or exotic), the spacetime solutions found here do not imply a violation of the energy conditions \cite{sam}, unlike the traversable wormholes.

The very existence of the spacetime-bridge solutions in vacuum gravity, as demonstrated here, appears to be an intriguing fact in itself. These may serve as useful test beds for ideas regarding topology change in classical gravity and causality.
Issues such as these, as well as those related to the propagation of material particles in such geometries
remain open and are left to future investigations.

\acknowledgments
Thanks are due to Joseph Samuel, Romesh Kaul and Sayan Kar for general discussions. I am also indebted to Debraj Choudhury, whom the figures in this article owe a lot to. 
This work is supported by the SERB, Department of Science and Technology, Government of India through the grant no. ECR/2016/000027.

\end{document}